\documentclass[final,5p,times,twocolumn]{elsarticle}
 \biboptions{comma,sort&compress}
\usepackage{graphicx}
\usepackage{epstopdf}
\usepackage{here}
\usepackage[dvips]{epsfig}
\def\beq{\begin{equation}}
\def\eeq{\end{equation}}
\def\bea{\begin{eqnarray}}
\def\eea{\end{eqnarray}}
\def\bq{\begin{quote}}
\def\eq{\end{quote}}

\def\nnb{\nonumber}
\def\ga{\left(}
\def\dr{\right)}

\def\nnb{\nonumber}
\def\la{\langle}
\def\ra{\rangle}
\def\nin{\noindent}
\def\ba{\vspace*{-0.2cm}\begin{array}}
\def\ea{\end{array}\vspace*{-0.2cm}}

\def\b{$\bullet~$}
\def\als{\alpha_s}

\def\gg2{ \la\alpha_s G^2 \ra}
\def\gg3{g^3f_{abc}\la G^aG^bG^c \ra}
\def\ggg4{\la\als^2G^4\ra}

\def\beq{\begin{equation}}
\def\enq{\end{equation}}
\def\beqa{\begin{eqnarray}}
\def\enqa{\end{eqnarray}}
\def\nnb{\nonumber}

\def\lb{\label}

\journal{Elsevier}

\begin{document}
\pagestyle{myheadings}
\begin{frontmatter}

\title{$1^{--}$ and $0^{++}$ heavy four-quark and molecule states in QCD}
 \author[label1,label2]{R.M. Albuquerque\corref{cor1}\fnref{fn1}}
  \cortext[cor1]{Speaker, PhD Student fellow, FAPESP CNPq-Brasil.}
\ead{rma@if.usp.br}
\address[label1]{Instituto de F\'{\i}sica, Universidade de S\~{a}o Paulo, 
C.P. 66318, 05389-970 S\~{a}o Paulo, SP, Brazil}
\address[label2]{Laboratoire
Particules et Univers de Montpellier, CNRS-IN2P3, 
Case 070, Place Eug\`ene
Bataillon, 34095 - Montpellier, France.}
 \author[label3]{F. Fanomezana\corref{cor2}}
\ead{fanfenos@yahoo.fr}
\address[label3]{Institute of High-Energy Physics of Madagascar (iHEP-MAD), University of Antananarivo, 
Madagascar}
 \author[label2]{S. Narison}
    \ead{snarison@yahoo.fr}

 \author[label3]{A. Rabemananjara\corref{cor2}}
\ead{achris\_01@yahoo.fr}

\begin{abstract}
\nin
We estimate the masses  of the $1^{--}$ heavy four-quark and molecule states 
by combining  exponential Laplace (LSR) and finite energy (FESR) sum rules 
known perturbatively to lowest order (LO) in $\alpha_s$  but including non-perturbative 
terms up to the complete dimension-six condensate contributions. 
We use double ratio of sum rules (DRSR) for determining the $SU(3)$ breakings terms.  
The $SU(3)$ mass-splittings of about (50 -- 110) MeV and the ones of about (250 -- 300) MeV 
between the lowest ground states and their 1st radial excitations are (almost) heavy-flavour 
independent. 
The mass predictions summarized in Table \ref{tab:res} are compared with the ones in 
the literature (when available) and with the three $Y_c(4260,~4360,~4660) $ and 
$Y_b(10890)$ $1^{--}$ experimental candidates. We conclude that the lowest observed 
state cannot be a {\it pure} $1^{--}$ four-quark nor a {\it pure} molecule but may result from 
their mixings. We extend the above analyzes to the $0^{++}$ four-quark and molecule states 
which are  about (0.5-1) GeV heavier than the corresponding $1^{--}$ states, while the 
splittings between the $0^{++}$ lowest ground state and the 1st radial excitation is about 
(300-500) MeV. We complete the analysis by estimating the decay constants of the 
$1^{--}$ and $0^{++}$ four-quark states.
Our predictions can be tested using some alternative non-perturbative approaches 
or/and at LHC$_b$ or some other hadron factories.
\end{abstract}
\end{frontmatter}
\section{Introduction and a short review on the $1^{++}$ channel}
\vspace{-0.2cm}
\nin
A large amount of exotic hadrons which differ from the ``standard" $\overline cc$ 
charmonium and $\overline bb$ bottomium radial excitation states have been recently 
discovered in $B$-factories through $J/\psi\pi^+\pi^-$ and $\Upsilon\pi^+\pi^-$ 
processes and have stimulated different theoretical interpretations.  Most of them 
have been assigned as four-quarks and/or molecule states \cite{SWANSON}.
In previous papers \cite{X1,X2}, some of us have studied, using exponential QCD 
spectral sum rules (QSSR) \cite{SVZ,SNB,RRY} and the double ratio of sum rules (DRSR) \cite{DRSR,SNGh,HBARYON,SNFORM}, the nature of the $X(3872)$ $1^{++}$ 
states found by Belle \cite{BELLEX} and confirmed by Babar \cite{BABARX}, CDF 
\cite{CDF} and D0 \cite{D0}.
If it is a $(cq)(\overline{cq})$ four-quark or $D-D^*$ molecule state, one finds \cite{X1}:
\beq \hspace{-0.5cm} 
{X_c}=(3925\pm 127)~{\rm MeV}~, ~{\rm with} ~~\sqrt{t_c}=(4.15\pm 0.03)~{\rm GeV}
\lb{eq:x1}
\eeq \vspace{-0.1cm}
corresponding to a $t_c$-value common solution of the exponential Laplace (LSR) 
and Finite Energy (FESR) sum rules. While in the $b$-meson channel, using 
$m_b=4.26$ GeV,  one finds\,\cite{X1}:
\beq \hspace{-0.4cm} 
X_b=(10144\pm 104)~\rm {MeV}~, ~{\rm with}~~\sqrt{t_c}=(10.4\pm 0.02)~{\rm GeV} .~
\eeq
By assuming that the mass of the radial excitation $X'_Q\approx \sqrt{t_c}$, one can also deduce the mass-splitting: 
\beq
 X'_c-X_c\simeq 225~{\rm MeV}\approx X'_b-X_b\simeq 256~{\rm MeV}~,
\eeq
which is much lower than the ones of ordinary $\overline cc$ and $\overline bb$ states:
\beq
{\psi}(2S)-{\psi}(1S)\simeq 590\approx {\Upsilon}(2S)-{\Upsilon}(1S)\simeq 560~{\rm MeV},
\eeq
suggesting a different dynamics for these exotic states.
\vspace{-0.2cm}

\section{QCD Analysis of the $1^{--}$ and $0^{++}$ channels}
\vspace{-0.2cm}
\nin
In the following, we extend the previous analysis to the case of the $1^{--}$ and 
$0^{++}$ channels and improve some existing  analysis from QCD (spectral) 
sum rules in the $1^{--}$ channel \cite{RAPHAEL1,RAPHAEL2}. The results will 
be compared with  the experimental $1^{--}$ candidate states: 
$Y(4260), Y(4360), Y(4660), Y_b(10890)$. These states 
have been seen by Babar \cite{BABARY} and Belle \cite{BELLEY,BELLEYb} and 
which decay into $J/\psi\pi^+\pi^-$ and $\Upsilon\pi^+\pi^-$ around the $\Upsilon(5S)$ mass.
These states cannot be identified with standard $\overline cc$ charmonium and $\overline bb$ bottomium radial excitations and have been assigned to be four-quark or molecule states or some threshold effects.

\subsection*{\b Interpolating currents}
\nin
We assume that the $Y$ states are described either by the lowest dimension  
(without derivative terms) four-quark and molecule $\overline D_sD^*_s$ vector 
currents $J^{\mu}$ given by: 
{\footnotesize
\begin{eqnarray}\hspace{-2cm}
\label{eq:curr}
J^\mu_{4q} &=& \frac{\epsilon_{abc} \epsilon_{dec}}{\sqrt{2}}  \Bigg\{ 
	\Big{[} \left( s^T_a {\cal C} \gamma_5 Q_b \right) 
	\left( \overline{s}_d \gamma^\mu \gamma_5 {\cal C} \overline{Q}^T_e \right) + 
	\left( s^T_a {\cal C} \gamma_5 \gamma^\mu Q_b \right) 
	\left( \overline{s}_d \gamma_5 {\cal C} \overline{Q}^T_e \right) \Big{]} \nnb\\
	&& +b  \Big{[} \left( s^T_a {\cal C} Q_b \right) 
	\left( \overline{s}_d \gamma^\mu {\cal C} \overline{Q}^T_e \right) + 
	\left( s^T_a {\cal C} \gamma^\mu Q_b \right) 
	\left( \overline{s}_d {\cal C} \overline{Q}^T_e \right) \Big{]} \Bigg\} \\
\hspace{-2cm}
J^\mu_{mol} &=& {1\over \sqrt{2}}\ga {g'\over\Lambda'^2}\dr^2_{\rm eff}\Big{[} 
	\ga\overline s\gamma^\mu Q\dr \ga \overline Q s\dr + 
	\ga\overline Q\gamma^\mu s\dr\ga\overline s Q\dr\Big{]}
\end{eqnarray}}
\vspace{-0.5cm}

\subsection*{\b QCD input parameters}
\nin
The QCD parameters are given in Table {\ref{tab:param}} and we shall work with 
the running light quark parameters \cite{FNR,TARRACH}.
{\scriptsize
\begin{table}[hbt] \vspace{-0.5cm}
\setlength{\tabcolsep}{0.2pc}
 \caption{
QCD input parameters. }
    {\small
\begin{tabular}{lll}
&\\
\hline
Parameters&Values& Ref.    \\
\hline
$\Lambda(n_f=4)$& $(324\pm 15)$ MeV &\cite{SNTAU,BNP,PDG}\\
$\Lambda(n_f=5)$& $(194\pm 10)$ MeV &\cite{SNTAU,BNP,PDG}\\
$\hat m_s$&$(0.114\pm0.021)$ GeV &\cite{SNB,PDG}\\
$m_c$&$(1.26\sim1.47)$ GeV &\cite{SNB,SNHmass,PDG,SNH10,IOFFE}\\
$m_b$&$(4.17\sim4.70)$ GeV &\cite{SNB,SNHmass,PDG,SNH10}\\
$\hat \mu_q$&$(263\pm 7)$ MeV&\cite{SNB}\\
$\kappa\equiv \la \overline ss\ra/\la\overline uu\ra$& $(0.74\pm 0.06)$&\cite{HBARYON}\\
$M_0^2$&$(0.8 \pm 0.2)$ GeV$^2$&\cite{JAMI2,HEID,SNhl}\\
$\la\alpha_s G^2\ra$& $(7\pm 2)\times 10^{-2}$ GeV$^4$&
\cite{SNTAU,LNT,SNI,fesr,YNDU,SNHeavy,BELL,SNH10,SNG}\\
$\la g^3  G^3\ra$& $(8.3\pm 1.0)$ GeV$^2\times\la\alpha_s G^2\ra$&
\cite{SNH10}\\
$\rho\equiv \la \overline qq \overline qq\ra/\la \overline qq\ra^2$&$(2\pm 1)$&\cite{SNTAU,LNT,JAMI2}\\
\hline
\end{tabular}
}
\label{tab:param}
\end{table}
}

\section{$1^{--}$  four-quark state mass $Y_{Qq}$ from QSSR}
\vspace{-0.3cm}
\nin
In the following, we shall estimate the mass of the $1^{--}$ four-quark state 
$(\overline{Qq}) (Qq)$,
hereafter denoted by $Y_{Qd}$. In so doing, we shall use the ratios of the Laplace 
(exponential) sum rule and of FESR:
\beq
{\cal R}^{LSR}_{Qd}(\tau)\simeq M_{Y_{Qd}}^2 \simeq {\cal R}_{Qd}^{FESR}~.
\eeq
\vspace{-0.5cm}

\subsection*{\b The $Y_{cd}$ mass from LSR and FESR for the case $b$=0}
\begin{figure}[t] 
\begin{center}
\centerline {\hspace*{-7cm} a) }\vspace{-0.6cm}
{\includegraphics[height=25mm]{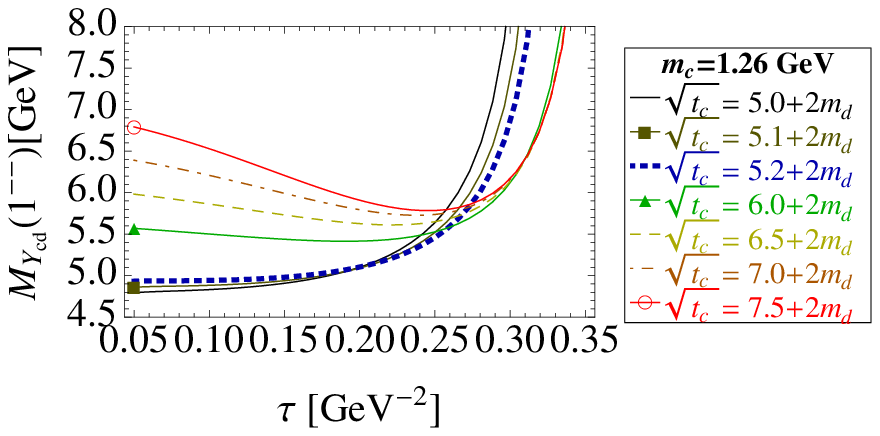}}
\centerline {\hspace*{-7cm} b) }\vspace{-0.6cm}
{\includegraphics[height=25mm]{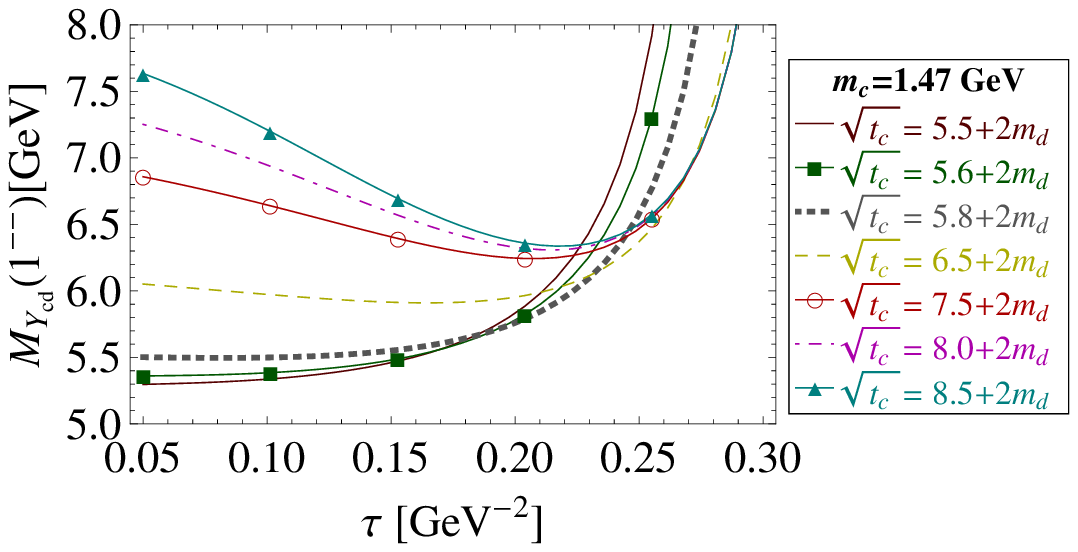}}
\centerline {\hspace*{-7cm} c) }\vspace{-0.6cm}
{\includegraphics[height=25mm]{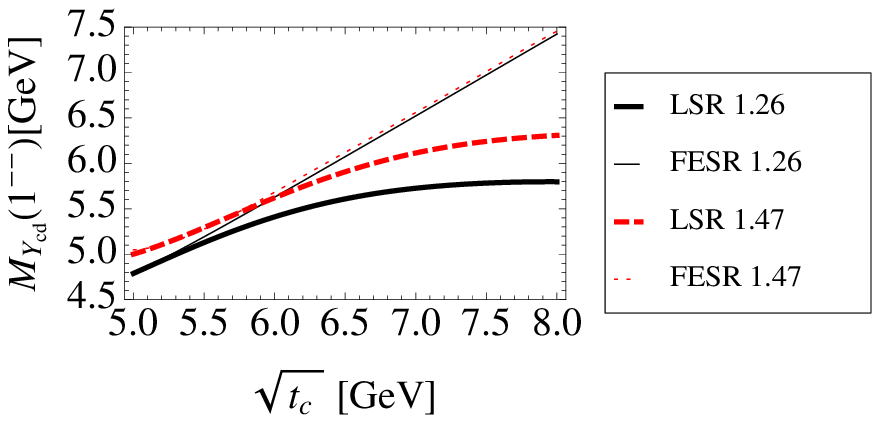}}
\caption{\scriptsize 
{\bf a)} $\tau$-behaviour of $M_{Y_{cd}}(1^{--})$ from ${\cal R}_{cd}^{LSR}$  for the 
current mixing parameter $b=0$, for different values of $t_c$ and for $m_c=1.26$ GeV;  
{\bf b)} The same as {\bf a)} but for $m_c=1.47$ GeV;  {\bf c)} $t_c$-behaviour of the LSR 
results obtained at the $\tau$-stability points and comparison with the ones from  
$ {\cal R}^{FESR}_{cd}$ for  $m_c=1.26$ and 1.47 GeV.}
\label{fig:yc} 
\end{center} \vspace{-0.6cm}
\end{figure}

\nin
Using the QCD inputs in Table \ref{tab:param}, we show the $\tau$-behaviour 
of $M_{Y_{cd}}$ from ${\cal R}^{LSR}_{cd}$ in Fig. \ref{fig:yc}a.
One can notice from Fig. \ref{fig:yc}a that the $\tau$-stability is obtained from 
$\sqrt{t_c}\geq 5.1$ GeV, while the $t_c$-stability is reached for $\sqrt{t_c}=7$ GeV. 
The most conservative prediction from the LSR is obtained in this range of 
$t_c$-values for $m_c=1.26$ GeV and gives in units of GeV:
\begin{eqnarray*}
4.79 \leq M_{Y_{cd}}\leq 5.73~~{\rm for}~~5.02 \leq \sqrt{t_c}\leq 7~{\rm and}~ m_c=1.26,\nnb\\
5.29 \leq M_{Y_{cd}}\leq 6.11~~{\rm for}~~5.5 \leq \sqrt{t_c}\leq 7~{\rm and}~m_c=1.47.
\end{eqnarray*}
We compare in Fig. \ref{fig:yc}b), the $t_c$-behaviour of the LSR results obtained at the 
$\tau$-stability points with the ones from  $ {\cal R}^{FESR}_{cd}$ for  
$m_c$=1.23 GeV (running) and 1.47 GeV (on-shell). One can deduce the common 
solution in units of GeV:
\bea
M_{Y_{cd}}&=& 4.814~~~~{\rm for}~~~~ \sqrt{t_c}=5.04(5)~~  {\rm and}~~m_c=1.26,\nnb\\
&=& 5.409~~~~{\rm for}~~~~ \sqrt{t_c}=5.6~~  {\rm and}~~m_c=1.47~.
\label{eq:ycd}
\eea
We observe that the on-shell $c$-quark mass value tends to overestimate 
$M_{J/\psi}$ \cite{X2,SNH10}. The same feature happens for the evaluation of the 
$X(1^{++})$ four-quark state mass \cite{X1}. 
Then, we are tempted to take as a final result in this paper the prediction  obtained by 
using the running mass $\overline{m}_c(m_c)=1262(17)$ MeV. 
Considering the uncertainties from the Table (\ref{tab:param}), we deduce
\bea
M_{Y_{cd}}
&=&4814(57) \mbox{ MeV}~.
\label{eq:ycd2}
\eea
Using the fact that the 1st FESR moment  gives a correlation between the mass of the 
lowest ground state and the onset of continuum threshold $t_c$, 
we shall approximately identify its value with the one of the radial excitation. 
Assuming that, one can deduce the mass-splitting:
$M'_{Y_{cd}}-M_{Y_{cd}}\approx 226~{\rm MeV},$
which is similar to the one obtained for the $X(1^{++})$ four-quark state \cite{X1}. 

\subsection*{\b The $Y_{bd}$ mass from LSR and FESR for the case $b=0$}
\nin
We extend the previous analysis to the $b$-quark sector, in order to estimate
the mass of $Y_{bd}$ four-quark state.
Considering, like in the case of charm, as a final estimate the one from the running 
$b$-quark mass $\overline{m}_b(m_b)=4177(11)$ MeV \cite{SNH10}, we deduce:
\bea
M_{Y_{bd}}
&=&11256(49) \mbox{ MeV}~.
\label{eq:ybd2}
\eea
From the previous result, one can deduce the value of the 
mass-splitting between the 1st radial excitation and the lowest mass ground state:
$M'_{Y_{bd}}-M_{Y_{bd}}\approx M'_{Y_{cd}}-M_{Y_{cd}}\approx 250~{\rm MeV},$
which are (almost) heavy-flavour independent and  also smaller than the one of the bottomium splitting 
($\sim \!560$ MeV).

\begin{figure}[tb] 
\begin{center}
a) {\includegraphics[height=25mm]{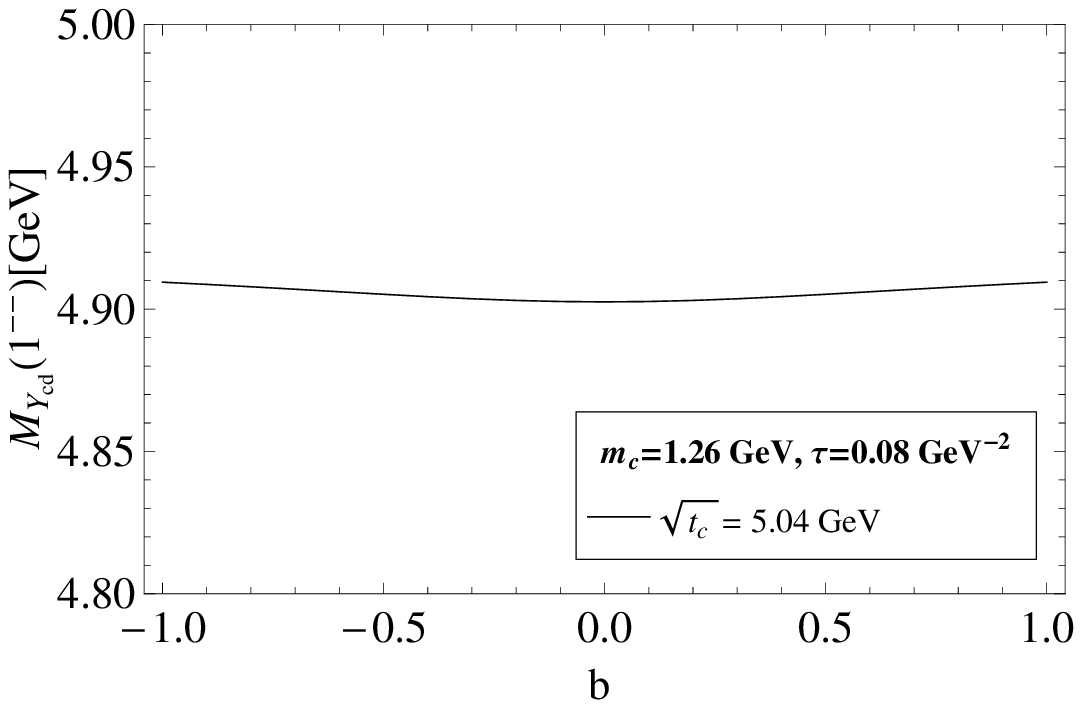}}
b) {\includegraphics[height=25mm]{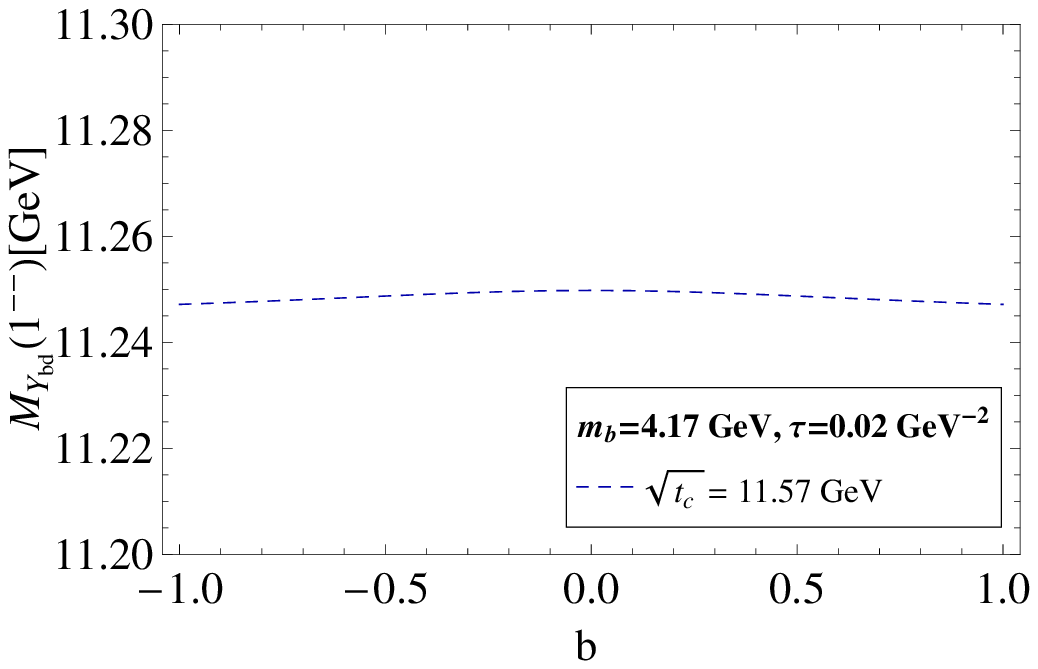}}
\caption{\scriptsize 
{\bf a)} $b$-behaviour of $M_{Y_{cd}}$  for given values of $\tau$ and $t_c$ 
and for $m_c=1.26$ GeV; {\bf b)} the same as a) but for $M_{Y_{bd}}$
and for $m_b=4.17$ GeV.}
\label{fig:b1} 
\end{center} \vspace{-0.8cm}
\end{figure}

In the following, we shall let the current mixing parameter $b$, as defined in Eq. (\ref{eq:curr}),
free and study its effect on the results obtained in Eqs. (\ref{eq:ycd2}) and (\ref{eq:ybd2}). In so 
doing, we fix the values of $\tau$ around the $\tau$-stability point and $t_c$ around the 
intersection point of the LSR and FESR. The results of the analysis are shown in Fig. \ref{fig:b1}.
We notice that the results are optimal at the value $b=0$.

For completing the analysis of the effect of $b$, we also study the decay constant $f_{Y_Qd}$
defined as:
$\la 0\vert j^\mu_{4q}\vert Y_{Qd}\ra=f_{Y_Qd}M^4_{Y_{Qd}}\epsilon^\mu~.$
Doing the analysis, giving  $M_{Y_{Qd}}$ and the corresponding 
$t_c$ obtained above, one can deduce the optimal values at $b=0$:
\beq
f_{Y_{cd}} \simeq 0.08~{\rm MeV}~~~~{\rm and} ~~~~~f_{Y_{bd}} \simeq 0.03~{\rm MeV}~,
\label{eq:f1}
\eeq
which are much smaller than $f_\pi=132$ MeV, $f_\rho\simeq 215$ MeV and 
$f_D\simeq f_B$=203 MeV \cite{SNFB}. One can also note that the decay constant 
decreases like $1/M_Q$ which can be tested in HQET or/and lattice QCD.

\subsection*{ \b $SU(3)$ breaking for $M_{Y_{Qs}}$ from DRSR}
\nin
We study the ratio $M_{Y_{Qs}}/M_{Y_{Qd}}$  using DRSR:
\beq
r^Q_{sd}\equiv {\sqrt{{\cal R}_{Qs}^{LSR}}/\sqrt{{\cal R}_{Qd}^{LSR}}}~~~~
{\rm where}~~~~Q\equiv c,~b~.
\eeq
Extracting the values at $\tau-$ and $t_c-$stabilities points we can deduce, 
respectively for $\sqrt{t_c}=5.1$ and 11.6 GeV:
\bea
r^c_{sd}&=&1.018(1)_{m_c}(5)_{m_s} (2)_{\kappa}(2)_{\overline uu}(1)_\rho~,\nnb\\
r^b_{sd}&=&1.007(0.5)_{m_b}(2)_{m_s} (0.5)_{\kappa}(1)_{\overline uu}(0.3)_\rho~.
\label{eq:ratioQs}
\eea
Using the results for  $Y_{Qd}$ in Eqs. (\ref{eq:ycd2}) and  (\ref{eq:ybd2}) and the
values of the $SU(3)$ breaking ratio in Eq. (\ref{eq:ratioQs}), we can deduce the 
mass of the $Y_{Qs}$ state in MeV:
\beq
M_{Y_{cs}}=4900(67)~,~~~~~~~~~~~
M_{Y_{bs}}=11334(55)~,
\eeq
leading to the $SU(3)$ mass-splitting:
$\Delta M^{Y_c}_{sd}\approx 87~{\rm MeV}\approx \Delta M^{Y_b}_{sd}\approx 78~{\rm MeV},$
which is also heavy-flavour independent. 

\section{ $1^{--}$ molecule masses from QSSR }
\vspace{-0.2cm}

\subsection*{\b The $\overline D^*_{d(s)}D_{d(s)}$ and $\overline B^*_{d(s)}B_{d(s)}$ molecules
\,\footnote{Hereafter, for simplicity $D$ and $B$ denote the scalar $D^*_0$ 
and $B^*_0$ mesons.}}
\nin
Like in the previous case, we use LSR and FESR for studying the masses of the 
$\overline D^*_dD_d$ and $\overline B^*_{d}B_{d}$ and DRSR for studying the $SU(3)$ 
breaking ratios:
\beq
r^D_{sd}\equiv  {M_{D^*_sD_s}/ M_{D^*_dD_d}}~,~~~~~~~~~~~~
r^B_{sd}\equiv  {M_{B^*_sB_s}/ M_{B^*_dB_d}}~.
\eeq
Using the sets ($m_c=1.26$ GeV, $\sqrt{t_c}=5.58~{\rm GeV}$) and 
($m_b=4.17~{\rm GeV}$, $\sqrt{t_c}=11.64(3)~{\rm GeV}$) common solutions of 
LSR and FESR, one can deduce in MeV:
\bea
M_{{D^*_dD_d}}
&=&5268(24) ~, ~~~~~~~~~
M_{{B^*_dB_d}}
=11302 (30) ~,\nnb\\
r^D_{sd}&=&1.018(1)_{m_c}(4){m_s}(0.8)_{\kappa}(0.5)_{\overline uu}(0.2)_{\rho}(0.1)_{G^3}~,\nnb\\
r^B_{sd}&=&1.006(1)_{m_b}(2){m_s}(1)_{\kappa}(0.5)_{\overline uu}(0.2)_{\rho}(0.1)_{G^3}~.
\label{eq:D*D}
\eea
Using the previous results in Eq. (\ref{eq:D*D}), one obtains in MeV :
\beq
M_{{D^*_sD_s}}=5363(33)~,~~~~~ M_{{B^*_sB_s}}=11370(40)~,
\label{eq:D*sDs}
\eeq
corresponding to a $SU(3)$ mass-splitting:
$\Delta M^{DD^*}_{sd}\simeq 95 ~{\rm MeV} \approx \Delta M^{BB^*}_{sd}\simeq 68~{\rm MeV}~.$

\subsection*{\b The  $J/\psi S_2$ and $\Upsilon S_2$ molecules}
\nin
Combining LSR and FESR, we consider the mass of the  $J/\psi S_2$ and 
$\Upsilon S_2$  molecules in a colour singlet combination, where 
$S_2\equiv \overline uu+\overline dd$ is a scalar meson
We work with the LO QCD expression obtained in \cite{RAPHAEL2}.
Then we analyze the $t_c$-behaviour of different $\tau$-extremas, from which 
we can deduce, for the running quark masses for $\sqrt{t_c}= 5.30(2)$ and 
10.23(3) GeV, in MeV: \vspace{-0.1cm}
\bea
M_{J/\psi S_2}
&=&5002(31) ~, ~~~~~ 
M_{\Upsilon S_2}
=10015(33) ~. 
\lb{eq:psis2}
\eea
The splitting (in units of MeV) with the first radial excitation approximately given by $\sqrt{t_c}$ is:
\vspace{-0.1cm}
\beq
M'_{J/\psi S_2}-M_{J/\psi S_2}\approx 298~,~~~~~~ M'_{\Upsilon S_2}-M_{\Upsilon S_2}\approx 213~.
\eeq
\vspace{-0.1cm}
In the same way, we analyse the $\tau$ and $t_c$ behaviours of the $SU(3)$ 
breaking ratios, from which, we can deduce:
\bea
r^\psi_{sd}&\equiv& {M_{J/\psi S_3}/ M_{J/\psi S_2}}=1.022(0.2)_{m_c}(5)_{m_s}(2)_{\kappa}~, \nnb\\
r^\Upsilon_{sd}&\equiv& {M_{\Upsilon S_3}/ M_{\Upsilon S_2}}=1.011(1)_{m_b}(2)_{m_s}(0.2)_{\kappa}~,
\label{eq:psisu3}
\eea
where $S_3\equiv \overline ss$ is a scalar meson. Then, we obtain in MeV:
\beq
M_{J/\psi S_3}=5112(41)~,~~~~~~~~~M_{\Upsilon S_3}=10125(40)~,
\lb{eq:psis3}
\eeq
corresponding to the $SU(3)$ mass-splittings:
$\Delta M^{J/\psi}_{sd}\simeq  \Delta M^{\Upsilon}_{sd}\approx 110~{\rm MeV}.$
Doing the same exercise for the octet current, we deduce the results in 
Table \ref{tab:res} where the molecule associated to the octet current is 
100 (resp. 250) MeV above the one of the singlet current for $J/\psi$ 
(resp $\Upsilon$) contrary to the $1^{++}$ case  discussed in \cite{X2}. 
The ratio of $SU(3)$ breakings are respectively 1.022(5) and 1.010(2) 
in the $c$ and $b$ channels which are comparable with the ones in 
Eq.(\ref{eq:psisu3}). \vspace{-0.4cm}

\section{$0^{++}$ four-quark and molecule masses from QSSR}
\vspace{-0.2cm}
\subsection*{\b $Y_{Qd}^0$ mass and decay constant from LSR and FESR}
\nin
We do the analysis of the $Y_{cd}^0$ and $Y_{bd}^0$ masses using 
LSR and FESR match. We work with the current mixing parameter $b=0$ from 
which we deduce in MeV, for the running quark masses, and respectively 
for $\sqrt{t_c}=6.5$ and 13.0 GeV:
\beq
M_{Y^0_{cd}}
=6125(51)~{\rm MeV}~,  ~~~
M_{Y^0_{bd}}
=12542(43)~{\rm MeV}~.~~
\eeq
One can notice that the splittings between the lowest ground state and 
the 1st radial excitation approximately given by $\sqrt{t_c}$ is:
$M'_{Y^0_{cd}}-M_{Y^0_{cd}}\approx 375 \:{\rm MeV}~,~~
M'_{Y^0_{bd}}-M_{Y^0_{bd}}\simeq 464 \:{\rm MeV}~,$
which is larger  than the ones of the $1^{--}$ states, comparable with the 
ones of the $J/\psi$ and $\Upsilon$, and are (almost) heavy-flavour 
independent. 
For completeness, we calculate the sum rule for the decay constants from which 
we deduce:
\vspace{-0.2cm}
\beq
f_{Y^0_{cd}} \simeq 0.12~{\rm MeV}~~~~{\rm and} ~~~~~f_{Y^0_{bd}} \simeq 0.03~{\rm MeV}~,
\label{eq:f0}
\eeq
which are comparable with the ones of the spin 1 case. 
\vspace{-0.3cm}

\subsection*{\b $SU(3)$ breaking for $M_{Y_{Qs} }^0$ from DRSR}
\vspace{-0.2cm}
\nin
Analyzing the $\tau$ and $t_c$ behaviours of the $SU(3)$, we deduce:
\bea
r^{0c}_{sd}&=&1.011(2)_{m_c}(3.8){m_s}(1.4)_{\kappa}(1)_{\overline uu}(0.7)_{\rho}~,\nnb\\
r^{0b}_{sd}&=&1.004(1)_{m_c}(1.7){m_s}(0.3)_{\kappa}~,
\eea
leading to the masses in MeV and theirs respective $SU(3)$ mass-splittings:
\vspace{-0.2cm}
\bea
M_{Y^0_{cs}} &=& 6192(59)~,  ~~~M_{Y^0_{bs}}= 12592(50)~, \nnb\\
\Delta M_{sd}^{Y^0_c} &\simeq& 67 ~\approx~ \Delta M_{sd}^{Y^0_b} ~\simeq~ 50~{\rm MeV}~.
\eea

\subsection*{\b $M_{D_{d}D_{d}}$ and $M_{B_{d}B_{d}}$ from LSR and FESR}
\nin
We consider as a final result the one corresponding to the 
running masses, for $\sqrt{t_c}=6.25(3)$ and 12.02 GeV:
\beq
M_{D_{d}D_{d}}
=5955(48) \mbox{ MeV}~, ~~~~
M_{B_{d}B_{d}}
=11750(40)  \mbox{ MeV} ~.
\eeq
The splittings (in MeV) between the lowest ground state and the 1st  radial excitation, 
approximately given by $\sqrt{t_c}$, is:
\beq
M'_{D_{d}D_{d}}-M_{D_{d}D_{d}}\approx 290~,~~~~~~M'_{B_{d}B_{d}}-M_{B_{d}B_{d}}\approx 270~,
\eeq
which, like in the case of the $1^{--}$ states are smaller than the ones of the $J/\psi$ and $\Upsilon$, and almost  heavy-flavour independent. 

\subsection*{\b $SU(3)$ breaking for $M_{\overline D_{s}D_{s}}$ and $M_{\overline B_{s}B_{s}}$ from DRSR}
\nin
Calculating the $SU(3)$ mass ratios for the molecules $\overline D_{s}D_{s}$ and 
$\overline B_{s}B_{s}$, considering different values of $t_c$ and the $t_c$ behaviour 
of their $\tau$-extremas, we deduce:
\begin{eqnarray*}
r^{0D}_{sd}&\equiv &{M_{D_sD_s}/ M_{D_dD_d}}=
1.015(1)_{m_c}(4){m_s}(2)_{\kappa}(1)_{\overline uu}(0.5)_{\rho}~,\nnb\\
r^{0B}_{sd}&\equiv& {M_{B_sB_s}/ M_{B_dB_d}}=
1.008(1)_{m_c}(4){m_s}(2)_{\kappa}(1)_{\overline uu}(0.5)_{\rho}~.
\end{eqnarray*}
Using the previous values of $M_{D_dD_d}$ and $M_{B_dB_d}$, we deduce:
\beq
M_{D_sD_s}=6044(56) \mbox{ MeV}~,~~~M_{B_sB_s}=11844(50) \mbox{ MeV},
\eeq
which corresponds to a $SU(3)$ splitting:
\beq
\Delta M^{DD}_{sd}\approx 89~{\rm MeV} \approx \Delta M^{BB}_{sd}\approx 94~{\rm MeV}~.
\eeq

{\scriptsize
\begin{table}[hbt]
\setlength{\tabcolsep}{0.7pc} \vspace{-0.6cm}
 \caption{
Masses of the four-quark and molecule states from the present analysis combining
Laplace (LSR) and Finite Energy (FESR). 
}
\vspace{-0.2cm}
    {\small
\begin{tabular}{llll}
&\\
\hline
States&& States &   \\
\hline
{\it Four-quarks}&\it $1^{--}$&&$0^{++}$\\
 $Y_{cd}$&4818(27)&$Y^0_{cd}$&6125(51) \\
$Y_{cs}$&4900(67)&$Y^0_{cs}$&6192(59) \\
$Y_{bd}$&11256(49)& $Y^0_{bd}$&12542(43)\\
$Y_{bs}$&11334(55)&$Y^0_{bs}$&12592(50) \\
{\it Molecules}&$1^{--}$&&$0^{++}$\\
$\overline D^*_dD_d$&5268(24)&$\overline D_dD_d$&5955(48)\\
$\overline D^*_sD_s$&5363(33)&$\overline D_sD_s$&6044(56)\\
$\overline B^*_dB_d$&11302(30)&$\overline B_dB_d$&11750(40)\\
$\overline B^*_sB_s$&11370(40)&$\overline B_sB_s$&11844(50)\\
{\it Singlet current}&$1^{--}$&{\it Octet current}&$1^{--}$\\
$J/\psi S_2$&5002(31)&&5118(29)\\
$J/\psi S_3$&5112(41)&&5231(40)\\
$\Upsilon S_2$&10015(33)&&10268(28)\\
$\Upsilon S_3$&10125(40)&&10371(45)\\
\hline
\end{tabular}
}
\label{tab:res}
\end{table}
}
\nin

\vspace{-0.7cm}
\section{Summary and conclusions}
\nin
\b The  three $Y_c(4260,~4360,~4660) $  $1^{--}$ experimental candidates are too low 
for being pure four-quark or/and molecule $\overline DD^*$ and $J/\psi S_2$ states but can 
result from their mixings.  The $Y_b(10890)$ is lower than the predicted values of the 
four-quark and $\overline BB^*$ molecule masses but heavier than the predicted $\Upsilon S_2$ 
and $\Upsilon S_3$ molecule states. Our results may indicate that some other exotic structure 
of these states are not excluded. 
\\
\b For the $1^{--}$, there is a regularity of about (250-300) MeV  for the value of  the 
mass-splittings between the lowest ground state and the 1st radial excitation roughly 
approximated by the value of the continuum threshold $\sqrt{t_c}$ at which the LSR and 
FESR match. These mass-splittings  are  (almost) flavour-independent and are much smaller 
than the ones of 500 MeV of ordinary charmonium and bottomium states.\\
\b  There is also a regularity of about 50--90 MeV for the $SU(3)$ mass-splittings of the 
different states which are also (almost) flavour-independent. \\
\b The spin 0  states are  much more heavier ($\geq$ 400 MeV) than the spin 1 states, like 
in the case of hybrid states \cite{SNB}. \\
\vspace*{-0.5cm}

\subsection*{\bf Acknowledgment}

\noindent
This work has been partly supported by the CNRS-IN2P3 
within the project Non-Perturbative QCD and Hadron Physics, 
by the CNRS-FAPESP program and by  CNPq-Brazil. 
S.N. has been partly supported
by the Abdus Salam ICTP-Trieste (Italy)  as an ICTP consultant for Madagascar.
\vspace*{ -.4cm}

\end{document}